\documentclass[fleqn,usenatbib]{mnras}

\usepackage{newtxtext,newtxmath}

\usepackage[T1]{fontenc}

\DeclareRobustCommand{\VAN}[3]{#2}
\let\VANthebibliography\thebibliography
\def\thebibliography{\DeclareRobustCommand{\VAN}[3]{##3}\VANthebibliography}

\usepackage{graphicx}	
\usepackage{amsmath}	

\usepackage[dvipsnames]{xcolor}

\usepackage{url}
\usepackage{multicol}
\defcitealias{burdanov2023gpu}{B23}

\title[Small body harvest with the ASTEP project]{Small body harvest with the Antarctic Search for Transiting Exoplanets (ASTEP) project}

\author[S. N. Hasler et al.]{
S. N. Hasler,$^{1}$\thanks{E-mail: shasler@mit.edu (SNH); burdanov@mit.edu (AYB)}
A. Y. Burdanov,$^{1}$\color{blue}\footnotemark[1]\color{black}
J. de Wit,$^{1}$
G. Dransfield,$^{2}$
L. Abe,$^{3}$
A. Agabi,$^{3}$
P. Bendjoya,$^{3}$\newauthor
N. Crouzet,$^{4}$
T. Guillot,$^{3}$
D. Mékarnia,$^{3}$
F.~X. Schmider,$^{3}$
O. Suarez,$^{3}$
A.H.M.J. Triaud$^{2}$
\\
$^{1}$Department of Earth, Atmospheric and Planetary Sciences, Massachusetts Institute of Technology, 77 Massachusetts Avenue, Cambridge, MA 02139, USA\\
$^{2}$School of Physics and Astronomy, University of Birmingham, Edgbaston, Birmingham B15 2TT, UK\\
$^{3}$Université  C\^ote d'Azur, Observatoire de la C\^ote d'Azur, CNRS, Laboratoire Lagrange, Bd de l'Observatoire, CS 34229, F-06304 Nice cedex 4, France\\
$^{4}$ European Space Agency (ESA), European Space Research and Technology Centre (ESTEC), Keplerlaan 1, 2201 AZ Noordwijk, The Netherlands 39
}

\date{Accepted 2023 September 22. Received 2023 September 21; in original form 2023 July 07}

\pubyear{2023}

\begin{document}
\label{firstpage}
\pagerange{\pageref{firstpage}--\pageref{lastpage}}
\maketitle

\begin{abstract}
Small Solar system bodies serve as pristine records that have been minimally altered since their formation. Their observations provide valuable information regarding the formation and evolution of our Solar system.  Interstellar objects (ISOs) can also provide insight on the formation of exoplanetary systems and planetary system evolution as a whole. In this work, we present the application of our framework to search for small Solar system bodies in exoplanet transit survey data collected by the Antarctic Search for Transiting ExoPlanets (ASTEP) project. We analysed data collected  during the Austral winter of 2021 by the ASTEP 400 telescope located at the Concordia Station, at Dome C, Antarctica. We identified 20 known objects from dynamical classes ranging from Inner Main-belt asteroids to one comet. Our search recovered known objects down to a magnitude of $V$ = 20.4\,mag, with a retrieval rate of $\sim$80\% for objects with $V \le $ 20\,mag. Future work will apply the pipeline to archival ASTEP data that observed fields for periods of longer than a few hours to treat them as deep-drilling datasets and reach fainter limiting magnitudes for slow-moving objects, on the order of $V\approx $ 23-24\,mag. 
\end{abstract}




\section{Introduction}

    Small Solar system bodies\footnote{See International Astronomical Union (IAU) Resolution B5 for definitions: \url{https://www.iau.org/static/resolutions/Resolution_GA26-5-6.pdf} (Accessed 2023 June 1)}, i.e. asteroids, comets, most Trans-Neptunian Objects (TNOs), and interplanetary dust particles are an interesting avenue of study in terms of understanding the Solar system's formation and evolution, as well as planetary impacts \citep{2014Natur.505..629D, michel2015asteroids, 2015SSRv..197..191D}. 
    Specifically, small bodies provide a unique opportunity to understand the environment of the early Solar System as these objects are the leftovers from the age of planet formation residing in the Solar System today. Additionally, interstellar objects (ISOs), which are not gravitationally bound to the Solar system, open up an entirely new avenue for placing the Solar system in a broader context and understanding planetary system formation more widely \citep{meech2017brief, guznik2020initial}.

    Most of our knowledge about different populations of small bodies comes from dedicated ground- and space-based surveys\footnote{\url{https://minorplanetcenter.net/iau/lists/MPDiscsNum.html} (Accessed 2023 June 11)}. Concurrent with these surveys, searches for small bodies have been performed on various archival data sets (e.g., \citealt{vaduvescu2009euronear, 2012PASP..124..579G,2020A&C....3000356V}). Indeed, many astronomical observations contain serendipitous detections of small bodies. Mining such data can be a useful tool for making statistical inferences about their population.

    
    This work highlights the extension of the work described in \citealt{burdanov2023gpu} (hereinafter \citetalias{burdanov2023gpu}), which developed a graphics processing unit (GPU)-based framework to uncover small Solar system bodies that cross the field of view (FoV) of a targeted exoplanet survey. The pipeline was designed as a wrapper for the synthetic tracking software, \textsc{Tycho Tracker} \citep{parrott2020tycho}, to search for small bodies and increase the scientific yield of the survey. It was applied to archival and last night data collected by the SPECULOOS (Searching for Planets EClipsing ULtra-cOOl Stars) ground-based photometric survey \citep{jehin2018speculoos, delrez2018ground, gillon2018searching}. In this work, we have adapted and applied the pipeline to one season of archival data from the Antarctic Search for Transiting ExoPlanets \citep[ASTEP; ][]{daban2010astep, guillot2015astep} project to assess the sensitivity of ASTEP for finding small bodies. 
    

    Antarctica has a much briefer history of astronomical observations compared to the rest of the world. The first optical astronomy research occurred in Antarctica in 1971 with a program to perform solar astronomy \citep{indermuehle2005history, pomerantz1981polar}, leading to a spectacular power spectrum and the first clear identification of the Sun's oscillations modes \citep{1980Natur.288..541G}. The cold, dry, and stable air there provides excellent observing conditions for astronomy \citep{crouzet2010astep, crouzet2018four}. An additional benefit is the nearly continuous three month long night during the Antarctic winter months, which allows for longer observations than can be obtained by standard mid-latitude surveys.
    
    ASTEP's location (latitude -75°06’01”.3, longitude 123°19’26”.5, elevation 3267\,m) makes it particularly unique for several other reasons, including that it has been the only photometric survey operating continuously for several years on the Antarctic continent. Other instruments include the Infrared Telescope Maffei also at Concordia \citep{2003MmSAI..74...37T, 2014SPIE.9145E..0DD}, the Antarctica Schmidt Telescopes (AST3) at the non-manned Kunlun station, Dome A \citep{yuan2014ast3} and the Multi-band Survey Telescope (MST) at the Zhongshan Station on the coast of Antarctica \citep{chen2023multi}. The ASTEP project began in 2006 and the first observations of the ASTEP 400 telescope began in 2010 \citep{daban2010astep, guillot2015astep, mekarnia2016transiting}. 
    
\begin{figure*}
        \centering
        \includegraphics[width=0.8\textwidth]{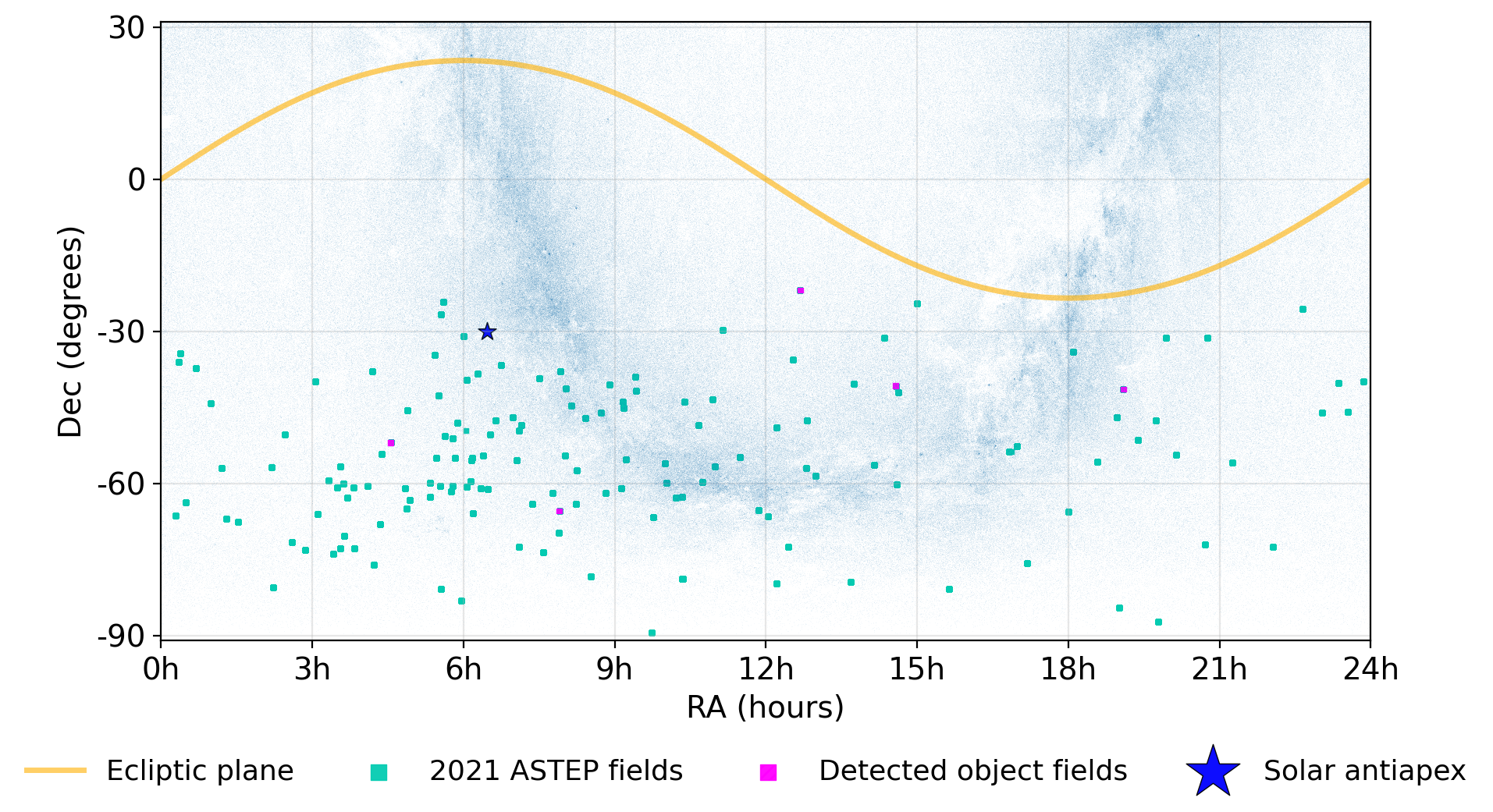}
        \caption{Projection of the celestial sphere with stars (blue dots) from the Tycho-2 catalog \citep{hog2000construction}, the ecliptic plane (orange line), the solar antiapex (blue star), and the processed ASTEP fields (teal squares) over-plotted. Fields with detections of known objects are also shown in magenta. The size of the markers depicts the size of ASTEP's field of view.}
        \label{fig:map}
    \end{figure*}

\begin{figure}
        \centering
        \includegraphics[width=0.9\columnwidth]{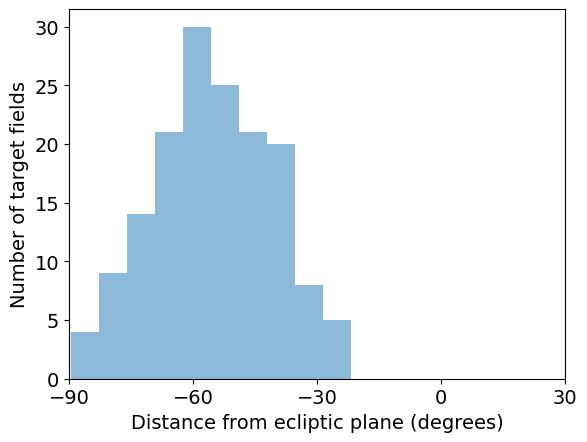}
        \caption{Distribution of observed target fields for the 2021 ASTEP 400 observations with respect to the ecliptic plane.}
        \label{fig:dec_distribution}
    \end{figure}

    ASTEP's vantage point from Antarctica is mostly away from the ecliptic, and also covers the region on the sky containing the solar antiapex (see Fig.~\ref{fig:map} and~\ref{fig:dec_distribution}). \citealt{hoover2022population} suggest that the ISO population is characterized by a clustering of trajectories in the direction of the solar apex and antiapex (the solar apex refers to the direction that the Sun travels with respect to the local standard of rest, with the antiapex directly opposite). ASTEP's point of view provides the potential to observe ISOs, should they enter the Solar system from the direction of the antiapex. Additionally, ASTEP's range of view may be uniquely suited to catch higher-inclination small bodies, whose evolutionary history are not yet well understood \citep{hromakina2021small}. Serendipitous detections of small bodies in the ASTEP data could thus provide better context for these groups if they happen to cross its FoV. 
    
    
    The rest of this paper is organized as follows, Section \ref{ASTEP_project} describes the ASTEP project and the data it provides. Section \ref{Data_processing} discusses the original data processing pipeline and changes made to process ASTEP data. In Section \ref{results}, we discuss the application of the pipeline to one year of ASTEP data. We summarise the findings of this work and discuss future applications in Section \ref{conclusions}.


\section{ASTEP Data}\label{ASTEP_project}

    The ASTEP program was designed to assess the ability to perform high-precision photometry to search for transiting exoplanets at Dome C, Antarctica \citep{crouzet2011astep} with ASTEP 400, a 40-cm telescope with a 1°x1° FoV \citep{daban2010astep, guillot2015astep}. Since, ASTEP has produced a number of observations and detections of difficult exoplanet transits, particularly of systems with long orbital periods and transit timing variations \citep[such as][]{Burt2021,Dransfield2022,Trifonov2023}. The unique location of the ASTEP observatories offers prime observing conditions for ground-based photometry, given an exceptionally clear sky, great seeing conditions, and near-uninterrupted observations during the three-month long Antarctic winter \citep{crouzet2010astep, crouzet2018four}. In 2022, ASTEP was moved to another astronomic dome on the Concordia station and upgraded to have the ability to perform simultaneous two-color photometry \citep{crouzet2020towards, dransfield2022observation,schmider2022observing}. 

    ASTEP 400 is a fully computer-controlled telescope. Observations are conducted in an automatic observation mode, or remotely operated when needed, from the beginning of March up to the end of September for each campaign. Data acquisition are performed when the Sun elevation is lower than $-$8°. During each campaign, more than 150 science fields are observed for a total of $\sim$ 2\,000 h of science exposure frames corresponding to about 5 TB of data. At the time of writing, the program has collected about 50 TB of data overall. The aim of this work is to begin maximizing the science return of this data by recovering serendipitous detections of small bodies that cross the FoV.

    The data acquired for this work comprises one season of observations from the ASTEP 400 telescope (MPC code P48), from March to October of 2021 (see Fig.~\ref{fig:map}). The data is comprised of 135,351 images from 167 target fields. The ASTEP target fields from 2021 consist primarily of \textit{TESS} (Transiting Exoplanet Survey Satellite) follow-up candidates. Images were taken with the FLI Proline science camera with 4k × 4k pixels and a pixel scale of 0.9 arcsec/pixel. A dichroic plate split the light for guiding and science, into $<$\,600\,nm and $>$\,600\,nm, respectively. Images were taken without a filter on the science camera, resulting in a bandpass of 600 - 800\,nm. This is close to the R Johnson-Cousins passband, and is thus designated as the ``R" filter. Image exposure times range from 2 to 200\,seconds.
    
    After removing defocused fields from the 2021 observations of bright targets (e.g., $\beta$ Pictoris and HD\,42933), we were left with 141 unique target fields for processing across 206 nights. This translates to 141 square degrees of sky coverage over the course of the observing season. The average observation duration per target was just over 4\,hours, with the longest observation duration being $\sim$13 hours. The data for the 206 nights were then passed through our data processing pipeline (see Section~\ref{Data_processing}).

\section{Data Processing Pipeline}\label{Data_processing}
    
    We developed a data processing and analysis pipeline, described in \citetalias{burdanov2023gpu}, to recover serendipitous detections of small bodies in a ground-based photometric survey. The pipeline is designed to employ a publicly-available, GPU-accelerated synthetic tracking software, \textsc{Tycho Tracker} \citep{parrott2020tycho}. The framework was initially tested and applied to data from the SPECULOOS survey, a targeted ground-based photometric survey searching for transiting planets around the nearest ultra-cool dwarf stars. The application of the framework to the SPECULOOS data and results are described in \citetalias{burdanov2023gpu}. 
    
    The pipeline structure is nearly identical to the structure described in \citetalias{burdanov2023gpu}, with minor changes made to account for the ASTEP data structure. The pipeline is coded in \textsc{Python} programming language, except the synthetic tracker, and run on a computer with a \textsc{Windows 10} operating system. The synthetic tracking portion happens in \textsc{Tycho Tracker} and uses \textsc{OpenCL} to communicate with the GPUs. A description of the pipeline and the updates made to process the ASTEP data is given in the following sections. 
    
\subsection{ASTEP Data Pipeline}
    The ASTEP data used for this work was provided as raw FITS files that must be calibrated and corrected accordingly before being passed to the synthetic tracker. Calibration was performed using standard bias and dark corrections. The ASTEP data do not provide flat-fields because it is not easy to gather twilight flats in their location and dome flats are not available, so we did not apply any flat-field correction during the calibration step of this work. We then perform additional correction steps to mask and remove bad pixels and defective columns from the science images. The pipeline also performs source extraction on the images for filtering. This process utilizes an image segmentation technique from \textsc{Photutils}, an \textsc{Astropy} affiliated \textsc{Python} package. More information on the removal of poor quality and defocused images can be found in \citetalias{burdanov2023gpu}. Affected images may include those that were taken in a defocused mode when ASTEP was observing a bright star (e.g., observations of the $\beta$ Pictoris target field).

    From mid-July to early September of 2021, an issue that prevented the shutter from closing on ASTEP rendered the bias and dark images unusable \citep{triaud_2023}. We separated 22 nights in the data from mid-July to early September that were affected by this issue by measuring the mean and standard deviation of the counts in the bias and dark images. These nights were processed by the pipeline without the bias and dark correction step. 
    
    Calibrated and filtered images are then checked for relatively even gaps in observation time between each image. After filtering, it is possible that bad images have been removed intermittently, causing uneven gaps in overhead. These uneven or large gaps in time significantly reduce the speed at which \textsc{Tycho Tracker} is able to process a field of observations. Large gaps in time mean there are more object search vectors that need to be considered. 
    If the ASTEP images have large gaps between observations, they are separated into groups at each gap to minimize computational time. 
    
    As with the original pipeline, calibrated and filtered images are then passed to the synthetic tracker in groups. 
    Images are aligned and processed by the synthetic tracker. After the shifting and stacking process, assuming objects are moving linearly on the FoV, \textsc{Tycho Tracker} returns parameters for any object candidates that may have been identified in the sequence of images. It outputs a set of tracks, or candidate detections, with information about their speed, position angle (PA), pixel coordinates, and the signal-to-noise ratio (S/N) of the detection. This information is cross-checked with the Minor Planet Center (MPC) database of small body orbits and the \textsc{Find\_Orb}\footnote{\url{https://www.projectpluto.com/find_orb.htm} (Accessed 31 May 2023)} software to report previously known objects. Coordinates of both known and unknown candidate objects are recorded and can then be submitted to the MPC after review. For further details on this process, see \citetalias{burdanov2023gpu}.

\subsection{Sensitivity Tests}

    To gather an understanding of how well the pipeline retrieved objects from the ASTEP data, we conducted detection efficiency tests using the NASA Jet Propulsion Laboratory (JPL) Small-Body Identification\footnote{\url{https://ssd-api.jpl.nasa.gov/doc/sb_ident.html} (Accessed 20 May 2023)} Application Program Interface (API). We compared our fields with detections to all of the known objects that were present in the FoVs to recover the retrieval rate of our pipeline. The results of our tests are described in detail in Section~\ref{results}.

    We also performed injection-retrieval tests on the fields with detected known objects. We injected 100 synthetic moving objects into each set of FITS files prior to calibration for nights with known object detections. The synthetic object injection module is described in detail in \citetalias{burdanov2023gpu}. Fields with injected objects were processed by the pipeline and the detected objects were compared with the injected ones. The results agree with the detection efficiency tests above and are described in Section~\ref{results}.

\section{Results and Discussion}\label{results}

    \begin{table*}
\begin{center}

    \caption{All processed target fields. Fields with known small body detections are listed first, followed by fields with  no detections. Latitude and longitude coordinates are in the geocentric mean ecliptic reference frame. Undetected known objects are listed down to the faintest $\mathrm{V_{mag}}$ detected in the target fields with known object detections. The full table is available online.} \label{table:detected_fields}

\begin{tabular}{lccccccccc}
\hline
Target name  & \begin{tabular}[c]{@{}c@{}}Ecliptic \\longitude \\(deg)\end{tabular} & \begin{tabular}[c]{@{}c@{}}Ecliptic \\latitude \\(deg)\end{tabular} & \begin{tabular}[c]{@{}c@{}}RA \\(hms)\end{tabular} & \begin{tabular}[c]{@{}c@{}}Dec \\(dms)\end{tabular} & Filter & \begin{tabular}[c]{@{}c@{}}Observed \\ hours\\ \end{tabular} & \begin{tabular}[c]{@{}c@{}}Detected \\ known objects \end{tabular} & \begin{tabular}[c]{@{}c@{}}Undetected \\ known objects \end{tabular} & \begin{tabular}[c]{@{}c@{}}Faintest \\ $\mathrm{V_{mag}}$ \end{tabular} \\ 

\hline

TOI-772  & 198.117 & --16.049 & 12 40 46 & --21 52 22 & $R$ & 15.3 & 14 & 11  & 20.4 \\

TOI-1955 & 43.157  & --71.902 & 04 33 40 & --51 57 22 & $R$ & 1.7 & 1 & 1 & 19.0 \\

TOI-823  & 229.541 & --24.235 & 14 34 38 & --40 44 23 & $R$ & 7.8 & 2 & 2 & 18.6 \\

TOI-283  & 187.709 & --78.424 & 07 54 17 & --65 26 29 & $R$ & 9.0 & 1 & 1 & 17.0 \\

TOI-1130 & 282.800 & --18.729 & 19 05 30 & --41 26 15 & $R$ & 7.7 & 2 & 4 & 18.8 \\

HD1397b  & 319.958 & --58.526 & 00 17 47 & --66 21 32 & $R$ & 4.4 & -- & -- & --  \\

HD37781  & 77.569  & --73.848 & 05 38 17 & --50 38 27 & $R$ & 4.2 & -- & -- & --   \\

TOI-1027 & 181.303 & --32.063 & 11 08 32 & --29 39 11 & $R$ & 6.8 & -- & -- & --   \\

TOI-1041 & 177.183 & --64.697 & 09 13 36 & --55 11 52 & $R$ & 1.0 & -- & -- & --    \\

TOI-1054 & 291.765 & --33.280 & 20 08 27 & --54 19 03 & $R$ & 2.1 & -- & -- & --    \\
\\                                            
\end{tabular}

\end{center}

\end{table*}
    \begin{table*}
\begin{center}

    \caption{Objects detected in each of the five target fields and their corresponding magnitudes, orbit groupings, and object numbers. Orbit grouping and absolute magnitude ($H$) for each object were retrieved from the JPL Horizons Database.}\label{table:detected_objects}

\begin{tabular}{lccccc}
\hline
Object name  & \begin{tabular}[c]{@{}c@{}}Object number\end{tabular} & \begin{tabular}[c]{@{}c@{}}Orbit grouping\end{tabular} & \begin{tabular}[c]{@{}c@{}}Visual magnitude\end{tabular} & \begin{tabular}[c]{@{}c@{}}Absolute magnitude\end{tabular} & \begin{tabular}[c]{@{}c@{}}Target field\\ \end{tabular} \\ 

\hline

2001 VJ46   &  194412 & MBA      & 19   &  15.61  & TOI-772 \\

2013 AZ119  &  542241 & MBA      & 20   &  17.67  & TOI-772 \\

2001 OK69   &  618481 & MBA      & 20   &  16.38  & TOI-772 \\ 

2010 TZ18   &  264237 & MBA      & 15.5 &  15.84  & TOI-772 \\ 

Tuchkova    &  3803   & MBA      & 15.1 &  11.33  & TOI-772 \\ 

2001 OD25   &  63471  & MBA      & 19.3 &  15.08  & TOI-772 \\ 

2012 DH7    &  454003 & MBA      & 20   &  16.65  & TOI-772 \\

2001 OA108  &  54941  & MBA      & 17.5 &  13.37  & TOI-772 \\ 

2000 RC88   &  62087  & MBA      & 18.3 &  14.44  & TOI-772 \\ 

2012 HO39   &  459337 & MBA      & 19.6 &  17.14  & TOI-772 \\ 

2012 JG36   &  542029 & MBA      & 19.8 &  16.30  & TOI-772 \\

2004 XG58   &  230948 & MBA      & 20.4 &  15.80  & TOI-772 \\

2004 XX97   &  198521 & MBA      & 20.2 &  15.85  & TOI-772 \\

2001 VV89   &  111121 & MBA      & 18   &  14.14  & TOI-772 \\

2005 EH243  &  299140 & Hungaria & 19   &  17.85  & TOI-1955 \\

1998 WK1    &  137063 & MBA      & 18.4 &  14.86  & TOI-823 \\ 

1999 TV264  &  237453 & MBA      & 18.6 &  14.34  & TOI-823 \\

2000 GL81   &  81393  & MBA      & 18.8 &  12.70  & TOI-1130 \\ 

1999 GN33   &  101788 & MBA      & 18.4 &  16.08  & TOI-1130 \\  

C/2019 T2   &        & Comet    & 18   &  13.92  & TOI-283 \\
                                          
\end{tabular}

\end{center}

\end{table*}

    We analysed observations from the ASTEP 400 telescope from 206 nights in 2021, with dates between 1 March 2021 to 5 October 2021, which included 165 total target fields, with 141 unique targets. 
    The average duration of observations was $\sim$4\,hours. 
    No new, unknown moving objects were discovered, but we detected previously known objects in 5 of the fields across 6 of the observing nights.
    All target fields from 2021 are shown in Figure~\ref{fig:map} and Table~\ref{table:detected_fields}. 
    Target fields with detected objects are also indicated in Table~\ref{table:detected_fields} and shown in magenta in Figure~\ref{fig:map}. 
    
\begin{figure}
        \centering
        \includegraphics[width=0.9\columnwidth]{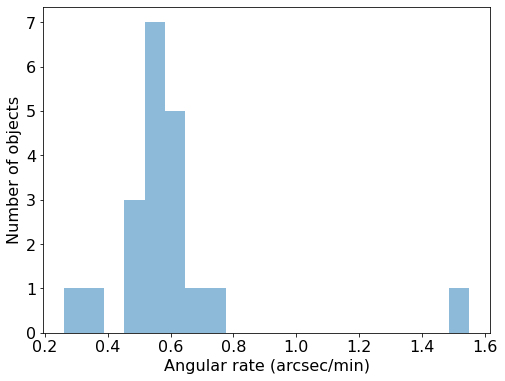}
        \caption{Total apparent angular rate in the plane-of-sky distribution for known objects detected in the 2021 ASTEP data.}
        \label{fig:detected_speed}
    \end{figure}

\begin{figure}
        \centering
        \includegraphics[width=0.9\columnwidth]{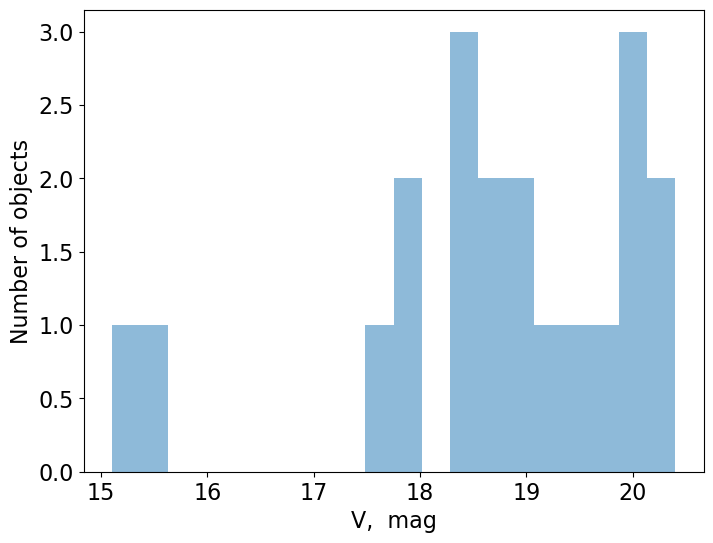}
        \caption{$V$\,mag distribution of known objects detected in the 2021 ASTEP data.}
        \label{fig:detected_mag}
    \end{figure}

    All detected objects are listed in Table~\ref{table:detected_objects}, along with their corresponding numbers, magnitudes, orbit groupings, and the fields in which they were recovered.
    The detected objects span several dynamical classes, including one inner main-belt (IMB) asteroid of the Hungaria group, 18 main-belt asteroids (MBAs), and one comet (see Table \ref{tab:orb_params} for dynamical class definitions). The dynamical classes of the bodies were retrieved from the JPL Horizons Database\footnote{\url{https://ssd-api.jpl.nasa.gov/doc/sbdb.html} (Accessed 31 May 2023)}. The speed and apparent angular rate on the plane-of-sky for each of these objects is presented Fig. \ref{fig:detected_speed}; their $V$\,mag distribution is presented in Fig. \ref{fig:detected_mag}. 

    The speeds of the detected objects range from the slowest moving detection at 0.26\,arcsec/min, to the fastest moving detection at 1.55\,arcsec/min. Generally, we need an object to move at least 2 pixels in order to be detected in our search. Given a pixel scale of 0.9 arcsec/pixel and the longest observing run of 13 hours for this data set, we can detect objects moving linearly as slow as 0.003 arcsec/min.
    
\begin{figure}
        \centering
        \includegraphics[width=0.4\textwidth]{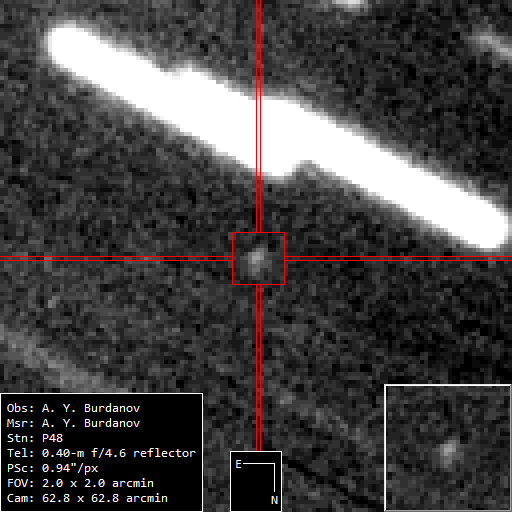}
        \caption{Detection of the faintest known object in the data. 2004 XG58 is an MBA detected with a magnitude of $V$ = 20.4\,mag and an angular rate of 0.5\,arcsec/min on 20 April 2021. The shifted and stacked image is composed of 50 images, each with an exposure time of 120\,seconds in the {TOI}-772 field.}
        \label{fig:2004XG58}
    \end{figure}

    The $V$\,mag of the detected objects ranges from 15.1\,mag for the brightest object and 20.4\,mag for the faintest. The median $V$\,mag of the detected objects is 18.9\,mag. The faintest detected object with a $V$\,mag of 20.4 is the MBA 2004 XG58. It was detected on 20 April 2021 with an angular rate of 0.5\,arcsec/min and position angle of 296.7\,deg. The shifted and stacked image is presented in Figure \ref{fig:2004XG58}. It is composed of 50 images with exposure times of 120\,seconds from the TOI-772 target field. 

        \begin{table}
        \centering
        \caption{The orbital elements of groups of small bodies as defined by the JPL Horizons Database, where $a$ is the semi-major axis (au), $q$ is the perihelion distance (au), and $Q$ is the aphelion distance (au).}
        \begin{tabular}{lc}
        \hline
            Orbital grouping & Orbital parameters \\
        \hline
            Inner Main-belt Asteroid & $a < 2.0, q > 1.666$ \\
            Main-belt Asteroid & $2.0 < a < 3.2, q > 1.666$ \\
            Comet-like & $Q > 5.0$ \\
        \end{tabular}

        \label{tab:orb_params}
    \end{table}
    
\begin{figure}
        \centering
        \includegraphics[width=\columnwidth]{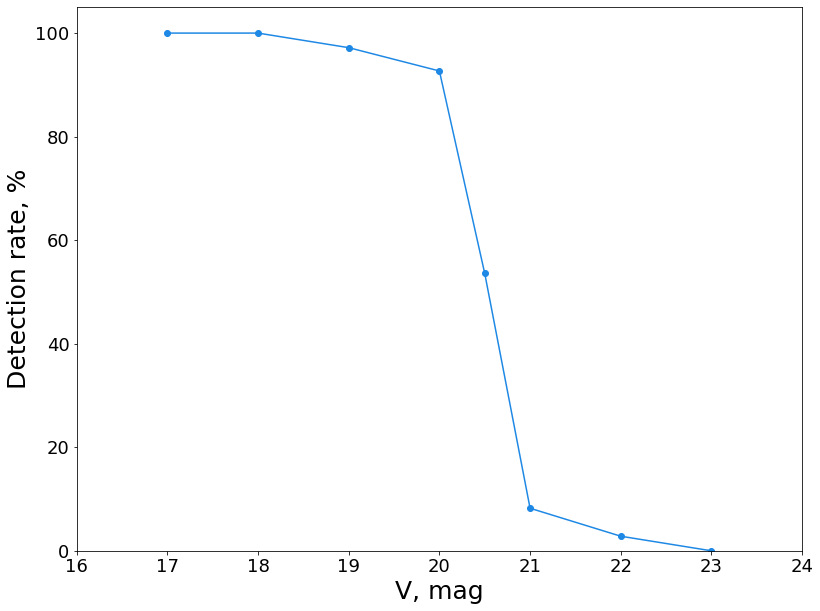}
        \caption{Injection-retrieval tests for slow-moving targets (0.2\,arcsec/min) performed on the data obtained in target field TOI-772 from 20 April 2021.}
        \label{fig:det_eff}
    \end{figure}

    To assess efficiency of the pipeline, we compared detected known small bodies with those predicted to be within an FoV at the time of observations, as well as injection-retrieval tests. We used the NASA JPL Small-Body Identification API to obtain small bodies' orbits (which were numerically integrated using a high-fidelity force model). According to the ephemerides, 39 small bodies were present in the FoVs with speeds in the 0.1-2.0\,arcsec/min range and brightness ranging from $V$=17.5\,mag to $V$=25.0\,mag (with a median value $V$=21.5\,mag). Our pipeline was able to detect 80\% of all possible known  objects down to $V$=20.0\,mag. No Near-Earth Asteroids (NEAs) were present in the FoVs. This result is in agreement with the injection-retrieval tests where objects with speed of 0.2\,arcsec/min were injected (see~Fig.~\ref{fig:det_eff}). Most of the undetected known objects were too faint or spent too little time with the FoVs to be detected. 

    We did not detect any unknown objects in our search. We attribute non-detections to the location of the target fields probed by the ASTEP project and its limiting magnitude. Most target fields are not located near the ecliptic, which is where the highest concentration of MBAs and Near-Earth objects is located \citep{2015aste.book..493M,2018ARA&A..56..137N}. As one moves away from the ecliptic, the detection probability decreases. The majority of the target fields in this survey are located far from the ecliptic, concentrating around 60 degrees in declination (see Fig.~\ref{fig:dec_distribution}). We expect the number of detected objects would be higher if the target fields were located nearer to the ecliptic (e.g., within 20 degrees). Another important factor contributing to the number of object detections is that the Southern sky (including high ecliptic latitudes) was extensively searched down to magnitude $\sim$\,20 by several surveys, such as Siding Spring Survey \citep{2003DPS....35.3604L} and ATLAS \citep{2018PASP..130f4505T}, while ASTEP's limiting magnitude is $\sim$\,20.4\,mag.

     In future work, we will apply the pipeline to archival data from the ASTEP project that was obtained during 2010 - 2013. The data presented in this paper was obtained primarily to confirm candidate transiting exoplanets identified by {\it TESS}, which resulted in an average observation time of $\sim$4\,hours per target and a maximum observation time of almost 13\,hours. Previous ASTEP observation strategies (e.g., observations from 2010-2013) observed target fields for periods of time longer than a few hours and over the course of many days \citep{mekarnia2016transiting}. Such extended observations will allow us to treat the data as deep-drilling sets to reach fainter limiting magnitudes than $V$ = 20.4\,mag and search for more slowly moving objects, similar to the archival data treatment in \citetalias{burdanov2023gpu}. For example, the WASP-19 field was observed by ASTEP for 24 nights during May 2010 \citep{abe2013secondary}. If we were to combine images from 150\,hours of observations over the course of several nights, we would have the potential to reach a limiting magnitude of $V$ = 23-24\,mag for slow-moving objects that remain in the FoV.

    \begin{figure*}
        \centering
        \includegraphics[width=0.8\textwidth]{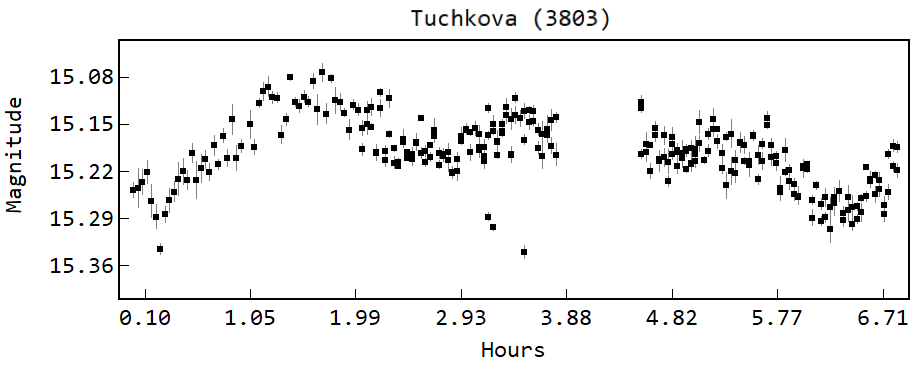}
        \caption{Light curve of a known object recovered in observations from 09 April 2021. The known object is main-belt asteroid Tuchkova  (3803). It has a median $V$\,mag of 15.5. The gap in observations around 4 hours corresponds to the object's passing in front of a background star.}
        \label{fig:lightcurve}
    \end{figure*}
    
    In terms of searching for interstellar objects (ISOs) in the data near the solar antiapex, the 2021 data included very few fields covering that region from 2021 (see Fig.~\ref{fig:map}). The known ISO ‘Oumuamua (1I/2017 U1) was detected with a magnitude of 19.8, at a heliocentric distance of 1.21 au, and with an on-sky speed of $\sim$15 arcsec/min\footnote{\label{iso_note1}\url{https://ssd.jpl.nasa.gov/horizons} (Accessed 31 May 2023)} \citep{bannister2017col, mpc_oumuamua}. 2I/Borisov (C/2019 Q4) was detected with a magnitude of 17.8, at a heliocentric distance of 2.84 au, and with an on-sky speed of 1.178 arcsec/min\footref{iso_note1} \citep{guznik2020initial, mpc_2IBorisov}.
    These parameters fall within the limitations of our search, meaning ASTEP has the capacity to recover similar ISOs should any cross the FoVs in additional datasets.

    We also constructed photometric light curves for the detected known objects. Light curves can provide detailed information about the physical properties and compositions of small bodies. \textsc{Tycho Tracker} maintains the capability to construct light curves of any detected objects. The light curve for the MBA Tuchkova (3803) is shown in Figure~\ref{fig:lightcurve}. Tuchkova is the brightest of the objects detected, with a magnitude of 15.1. The light curves of the remaining object detections be found in Appendix~\ref{appendix} and at the Asteroid Lightcurve Data Exchange Format (ALCDEF) database \citep{LCDB_warner}.
    
    The observations of Tuchkova span just under 7 hours, with an approximately 45 minute gap in the photometry around hour 4 of the observations (see Fig.~\ref{fig:lightcurve}) due to contamination by a background star. \cite{durech2020asteroid} report Tuchkova to have a sidereal rotation period of 6.07014 hours (Asteroid Lightcurve Database, LCBD; \citealt{LCDB_warner}). Our observations cover an entire rotation period for Tuchkova, minus the background star contamination.
    
    \textsc{Tycho Tracker} also offers the capability to determine the rotation periods of objects. We used \textsc{Tycho Tracker}'s built-in period search function to find the period of Tuchkova from our photometric measurements. The period search function works by finding the period that minimizes the sum of squared residuals from a curve computed as a least squares fit of a Fourier series. The period search returns a list of possible candidate periods each with their corresponding root mean square error (RMSE), and we select the candidate period of 6.158 hours, which has the smallest RMSE. 

    The rotation period reported by \citep{durech2020asteroid} was obtained from a model that was generated using 478 photometry measurements gathered by ATLAS \citep{2018PASP..130f4505T} across 2016-2018. While the period we recovered with \textsc{Tycho Tracker}'s built-in period search is close to the value reported in the literature, our observations only cover $\sim$7 hours of observations on one night. Ideally, we would have more observations that cover multiple rotation periods to gather a more robust estimate on the rotation period of the object.

    We see such densely sampled light curves as one of the main advantages of the ASTEP data over other surveys capable of observing similar parts of the sky. For instance, the Legacy Survey of Space and Time (LSST) at the Vera C. Rubin Observatory \citep{ivezic2019lsst} is expected to detect over 5 million new Solar system objects fainter than a magnitude of 16. However, light curves available from LSST will have variable sampling and most of them will be sparse \citep{2018arXiv180201783S, 2023ApJS..266...22S}, making light curve inversion (a technique used to model the surfaces of rotating objects from their brightness variations) more challenging.

\section{Conclusions}\label{conclusions}

    We have presented the application of our pipeline originally developed in \citetalias{burdanov2023gpu} to search for small bodies in data collected by the ASTEP project. The pipeline performs data processing and serves as a wrapper for the synthetic tracking software, \textsc{Tycho Tracker}. We analysed one season's worth of data collected by the ASTEP400 telescope during the 2021 Austral winter, in an effort to begin maximizing the science return of the survey. 

    In total, 167 target fields with 143 unique targets were observed by the ASTEP400 telescope in 2021, over the course of 216 nights. Of these fields, we processed 206 nights with 141 unique targets after removing defocused target fields. We identified 20 known objects in the data set in 5 fields across 6 nights of observations. The known objects span a few dynamical classes, including 1 inner main-belt asteroid, 18 main-belt asteroids, and 1 comet. Most of the objects were identified in fields near the ecliptic plane, which aligns well with the inclination distribution of small bodies in the Solar system. Object detections were made with no prior assumptions on objects' speed or position angles. After reviewing the candidate detections, we reported them to the MPC. The magnitudes of object detections ranged from $V$ = 15.1\,mag to $V$ = 20.4\,mag, and the objects' speeds ranged from 0.26\,arcsec/min to 1.55\,arcsec/min. We compared detected objects with those known to be present in the FoVs. The objects that were present but undetected were too faint to be recovered, with $V >$ 20.0\,mag. 
    
    Next steps will include applying the pipeline to archival data from the ASTEP project that was obtained during 2010 -- 2013. These observations were conducted for periods of time longer than a few hours and over the course of many days \citep{mekarnia2016transiting}, unlike the \textit{TESS} follow-up work. Extended observations will allow us to reach fainter limiting magnitudes than $V$ = 20.4\,mag and search for more slowly moving objects.
    
    Our data processing and analysis pipeline can continue to be adapted and applied to other photometric surveys to maximize their science return. The pipeline continues to reduce SPECULOOS data at the time of writing and has the potential to run as a "last night" process for ASTEP provided there are adequate computing resources at Concordia Station. The reduction of observations taken since December of 2021 would also allow us to obtain color information for any detected small body that crosses the FoV due to the new addition of a camera capable of two-color photometry on the ASTEP telescope \citep{schmider2022observing}.

\section*{Acknowledgements}

    We thank the reviewer, Dr. Meg Schwamb, for her time, attention, and constructive criticism. Dr. Schwamb's comments helped us to improve the quality of the paper. We thank and acknowledge the vital assistance of the French and Italian polar agencies (IPEV and PNRA), and all their staff, particularly the wintering-over staff of Concordia Station, without whom operations of ASTEP would not be possible. The authors would also like to thank Daniel Parrott, the developer of \textsc{Tycho Tracker}, for his advice and assistance with the software.  

    This research has made use of data and/or services provided by the International Astronomical Union's Minor Planet Center. 

    \noindent \textit{Software}: Tycho Tracker (\url{www.tycho-tracker.com}, \citealt{parrott2020tycho}), Astropy \citep{2013A&A...558A..33A, 2018AJ....156..123A}, {NumPy} \citep{harris2020array}, matplotlib \citep{Hunter:2007}. This research made use of Photutils, an Astropy package for detection and photometry of astronomical sources \citep{larry_bradley_2023_7946442}.

    This work has been supported by the NVIDIA Academic Hardware Grant Program. This research is supported from the European Research Council (ERC) under the European Union's Horizon 2020 research and innovation programme (grant agreement n$^\circ$ 803193/BEBOP) and from the Science and Technology Facilities Council (STFC; grant n$^\circ$ ST/S00193X/1).
 
\section*{Data Availability}

Photometry of detected objects is available at the Asteroid Lightcurve Data Exchange Format (ALCDEF) database. Objects' photometry was uploaded on 31 August 2023 and can be accessed by any user by either downloading the entire ALCDEF table or by performing an object search in the database. The remaining data is available on request to the corresponding author.



\bibliographystyle{mnras}
\bibliography{references} 



\appendix
\section{Light curves of detected objects}\label{appendix}
\begin{figure*}
    \begin{multicols}{3}
        \includegraphics[width=0.3\textwidth]{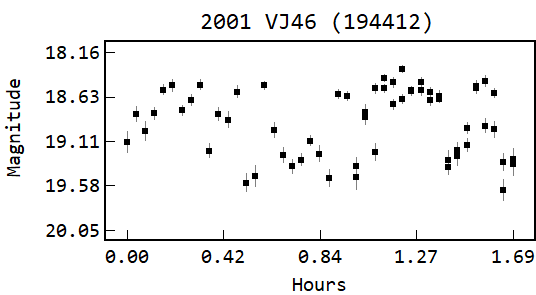}\par
        \includegraphics[width=0.3\textwidth]{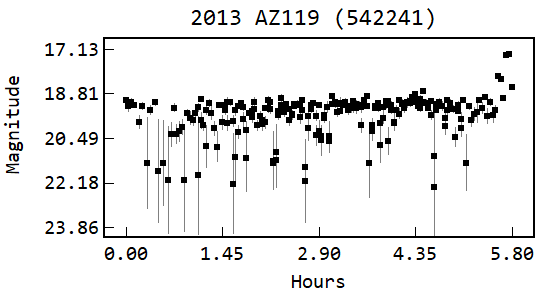}\par
        \includegraphics[width=0.3\textwidth]{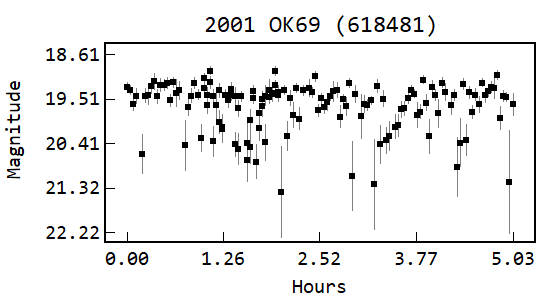}\par

        \includegraphics[width=0.3\textwidth]{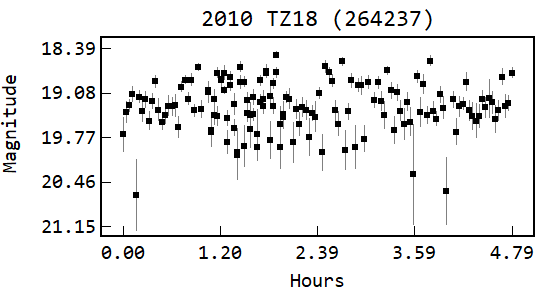}\par
        \includegraphics[width=0.3\textwidth]{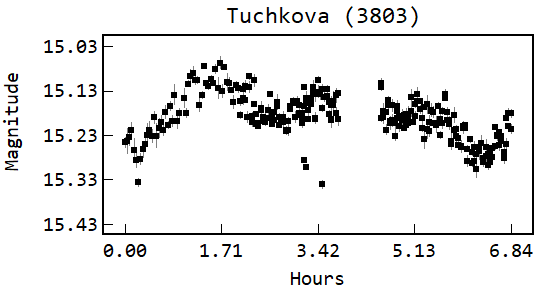}\par
        \includegraphics[width=0.3\textwidth]{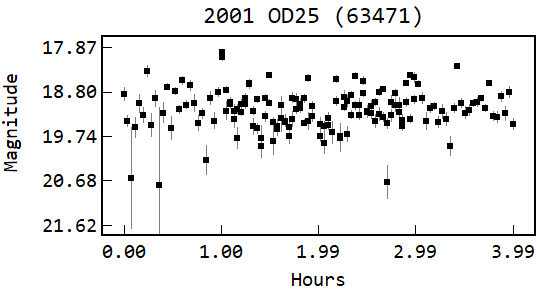}\par

        \includegraphics[width=0.3\textwidth]{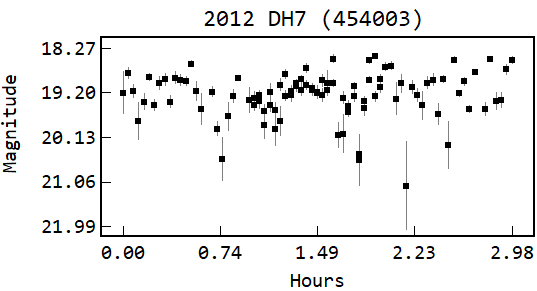}\par
        \includegraphics[width=0.3\textwidth]{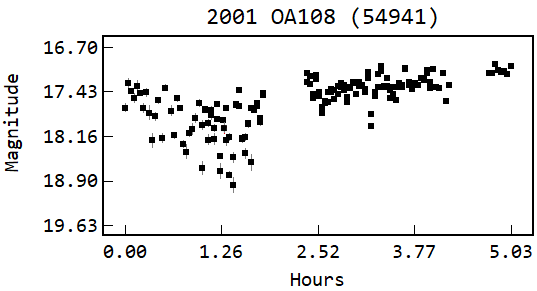}\par
        \includegraphics[width=0.3\textwidth]{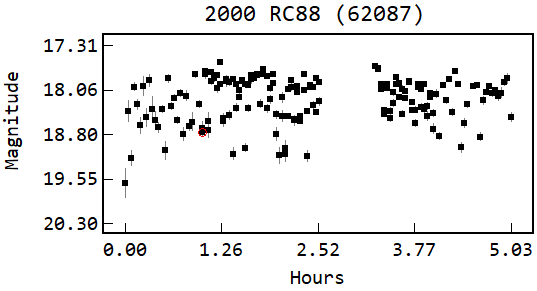}\par

        \includegraphics[width=0.3\textwidth]{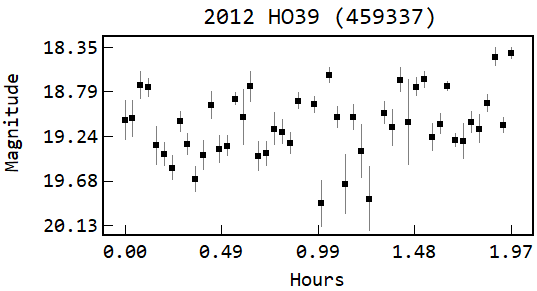}\par
        \includegraphics[width=0.3\textwidth]{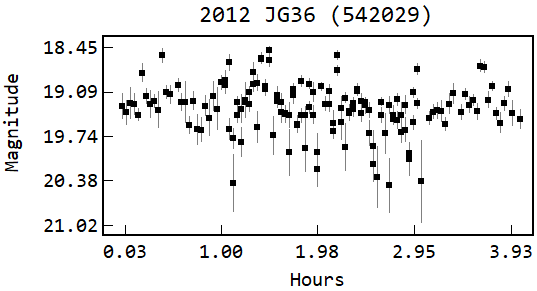}\par
        \includegraphics[width=0.3\textwidth]{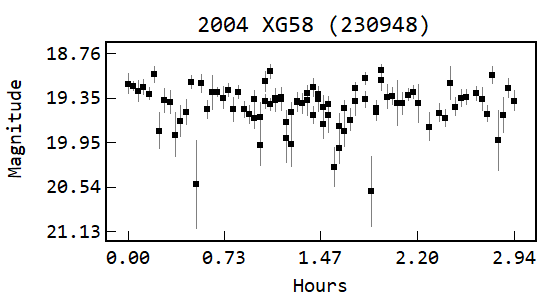}\par

        \includegraphics[width=0.3\textwidth]{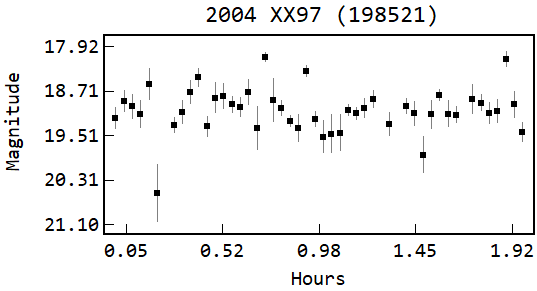}\par
        \includegraphics[width=0.3\textwidth]{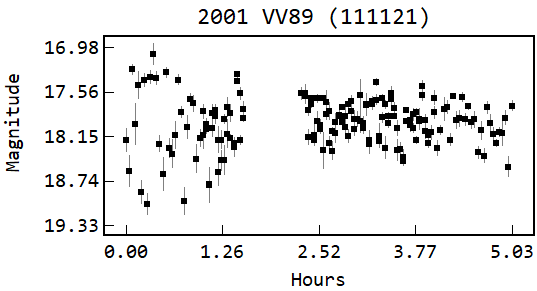}\par
        \includegraphics[width=0.3\textwidth]{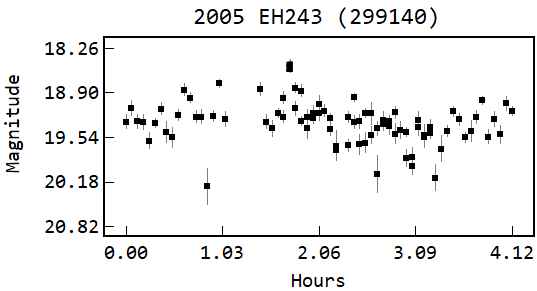}\par

        \includegraphics[width=0.3\textwidth]{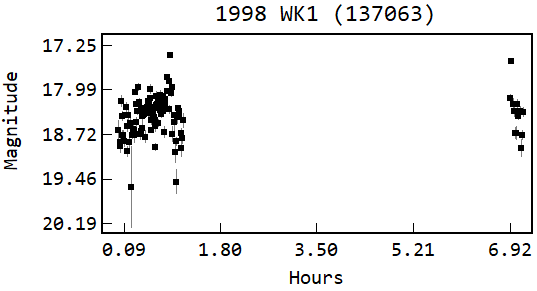}\par
        \includegraphics[width=0.3\textwidth]{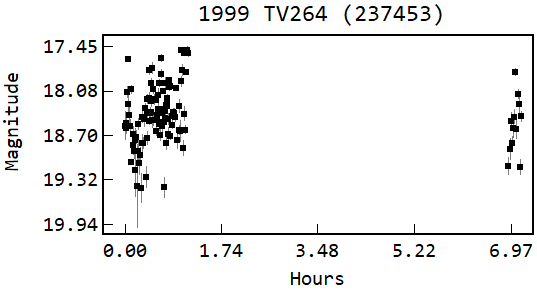}\par
        \includegraphics[width=0.3\textwidth]{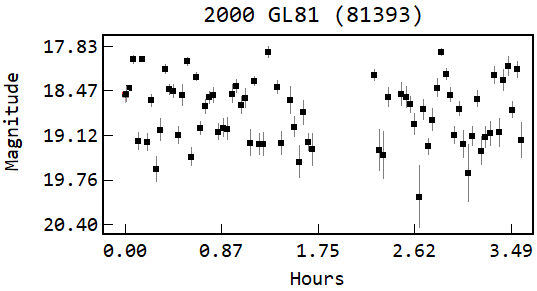}\par
    \end{multicols}
    \begin{multicols}{2}
        \includegraphics[width=0.3\textwidth]{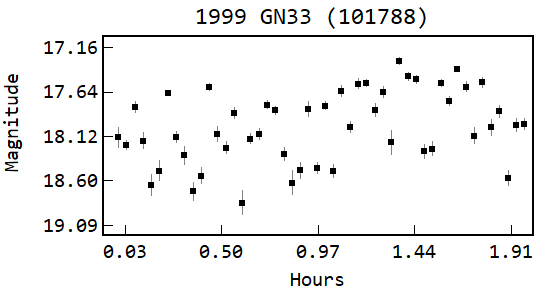}\par
        \includegraphics[width=0.3\textwidth]{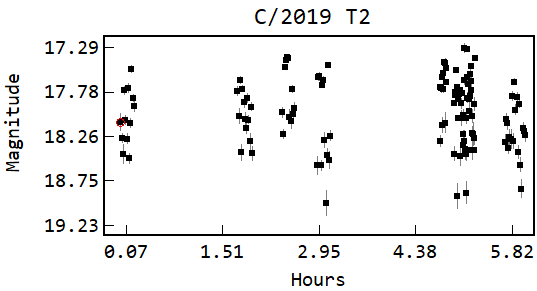}\par
    \end{multicols}
\caption{Light curves of all detected objects from the 2021 ASTEP data. Magnitude refers to apparent magnitude of the objects.}
\end{figure*}

\bsp	
\label{lastpage}
\end{document}